# Galmatheia: A Galactic Plasma Explorer

J. Edelstein, W. V. Dixon, and E. Korpela

*Space Sciences Laboratory, University of California, Berkeley, CA 94720*

Galmatheia is a broad bandpass (900–1800 Å), far-ultra violet (FUV) nebular spectrograph ($l/dl \sim 650$) for the study of the evolution of galactic plasma with a temperature of $10^{4.5} - 10^6$ K. Galmatheia will survey the FUV sky with 5' imaging and conduct hundreds of deep 8° x 5' field pointings during its proposed two-year mission. Unprecedented sensitivity is achieved by careful exclusion of FUV-bright stars and airglow background. The emission-line sensitivity for a single-day exposure and for a one-year sky survey 3° x 3° bin yields 50 $\sigma$ and 10–15 $\sigma$ detections, respectively, of both the predicted radiation from hot Galactic gas and previously-observed diffuse FUV emission. The continuum sensitivity provides 15–25 $\sigma$ detections of the predicted flux from unresolved extra-galactic sources.

## Measuring The Multi-Phase ISM

Models of plasma mixing and cooling are the key to our understanding of the structure and evolution of galaxies. Plasma evolution is a central element of physical processes acting on local to universal scales, such as galactic bubbles, super-bubbles, fountains and halos; galactic clusters; and the putative warm/hot intergalactic media (IGM). Very little is known about the morphology and distribution of diffuse, hot ($10^{4.5} - 10^6$ K) gas in both the Galactic disk and halo. Several models for the origin of this gas have been proposed. Each makes unique, verifiable predictions, yet none can be ruled out on the basis of current observations.

Galmatheia, an FUV spectrograph, will be used to measure the flow and balance of energy in the interstellar (IS) plasma by observing the primary cooling radiation from energetic, rapidly cooling gas. The spectrograph bandpass (900 - 1175 Å and 1335 - 1750 Å) includes the primary hot-plasma emission lines from both depletable and non-depletable elements, and its spectral resolution (~ 1.5Å and 2.3Å, respectively) allows the simultaneous observation and separation of the many diagnostic lines required to disentangle the effects of abundance and ionization equilibrium on plasma processes. The broad bandpass assures that a wide range of plasma states can be characterized without *a priori* assumptions of as yet unknown astrophysical conditions. The large field of view (8° x 5') and imaging resolution (5') are well matched to the study of galactic interstellar phenomena.

Galmatheia measurements will also reveal important details of $H_2$ fluorescence and dust scattering of starlight. A basic understanding of these processes is required to disentangle their emission from that of extragalactic systems, star-forming galaxies, or intergalactic sources.

## The Importance of FUV Diagnostics

The key diagnostics for both collisional and photoionized IS plasmas with temperatures $10^{4.5}$ to $10^6$ K are provided by cooling radiation in the FUV bandpass. The primary cooling lines of abundant elements in a variety of ionization states (e.g., C III 977 Å, O VI 1035 Å, C IV 1550 Å, and He II 1640 Å) are predicted to occur in the FUV for a host of galactic plasma cooling models.

There have been few sensitive attempts to measure diffuse FUV emission (Edelstein and Bowyer 1993, Korpela, Bowyer and Edelstein 1998), only a limited number of sight lines where FUV emission lines have been detected (Martin and Bowyer 1990, Dixon, Davidsen and Ferguson 1996), and no simultaneous C IV and O VI emission observations. (The C IV/O VI ratio is a key diagnostic for many galactic models.) Observatories that are optimized for point sources are not capable of sensitive spectral measurements of nebular IS emission over the large angular scales needed to characterize galactic plasma distributions (see Table 1).

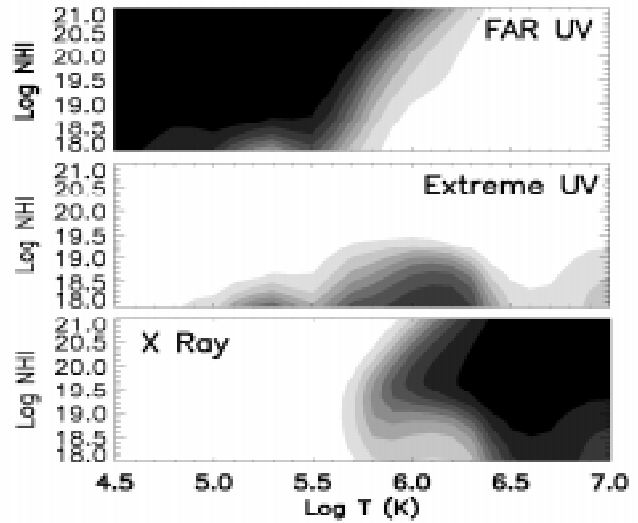

**Fig.1** *Fraction of energy observable for a hot equilibrium plasma as a function of log(T) and absorbing column. Black denotes >90 % fractional energy. Galmatheia FUV measures test gas undergoing phase transitions, wih $T = 10^{4.5} - 10^6$ K (top panel). EUV instruments (CHIPS, EUVE, EURD) can only measure the local ISM with $T \sim 10^6$ K gas (center panel). SXR instruments (ROSAT, DXS) can sense gas with $T >= 10^6$ K (bottom panel).*

| MISSION | FOV | λλ Å | δλ Å | AΩ cm² sr | AΩ_eff 1550 Å | AΩ_eff 1050Å |
|---|---|---|---|---|---|---|
| HST FOS | 4.3"x1.4" | 1175-3000 | 12 | 6.4x10⁻⁶ | 1.8x10⁻⁸ | |
| HST STIS | 51" X 2" | 1150-2300 | 3 | 1.1X10⁻⁴ | 1.8x10⁻⁶ | |
| HST COS | 2" φ | 1150-2400 | 1.5 | | 1.3x10⁻⁷ | |
| HUT / Astro 2 | 197"x19" | 800-1800 | 7 | 5.6x10⁻⁴ | | 2x10⁻⁶ |
| ORFEUS II | 26" φ | 900-1200 | 0.8 | 9.8x10⁻⁵ | | 1x10⁻⁷ |
| FUSE | 30"x30" | 900-1200 | 0.3 | 1.1x10⁻⁴ | | 1.5x10⁻⁶ |
| Voyager II | 0.87°x0.1° | 500-1100 | 30 | 5.6 x10⁻⁴ | | 3x10⁻⁷ |
| UVX | 4°x0.1° | 1350-1900 | 15 | 2.1x10⁻² | 1.2x10⁻⁴ | |
| DUVE FUCR | 4°x 0.12° | 980-1080 | 3.5 | 7.4x10⁻³ | | 6x10⁻⁶ |
| EURD | 26°x 8.0° | 350-1050 | 5 | 5.8x10⁻³ | | 8x10⁻⁶ |
| CHIPS | 26°x 16° | 90 - 260 | 2 | (EUV only) | | |
| GALMATHEIA | 8°x 0.08° | 950-1150 | 1.5 | 9.6x10⁻³ | | 1.4x10⁻⁴ |
| GALMATHEIA | 8°x 0.08° | 1350-1800 | 2.3 | 9.6x10⁻³ | 1.4x10⁻⁴ | |

*Table I Past and Future FUV Missions: The shaded section shows observatories optimized for point sources. Listed is field of view (FOV), bandpass ll, resolution dl, geometric grasp AW, and effective grasp AW_eff at 1550 Å and 1030 Å (With thanks to J. Dehavang.)*

The fraction of equilibrium thin-plasma radiative power in each of the FUV (900 - 1800 Å), Extreme-UV (EUV; 100 - 900 Å), and Soft X-ray (SXR; 0.1 - 10 keV) bandpasses is shown in Fig. 1 as a function of plasma temperature and intervening neutral absorbing column of gas. (We neglect the relatively small effects of dust absorption.) The FUV contains essentially all of the observable radiation for plasma with $N(HI) \geq 10^{18.5}$ cm⁻², such as gas in the galactic disk and halo, and $T \leq 10^6$ K, characteristic of the most rapidly-cooling IS phases wherein substantial radiative energy loss is anticipated to occur. SXR data (e.g., Snowden et al. 1997) are ill-suited for measuring plasmas with temperatures $T \leq 10^6$ K. In contrast to X-ray measurements, FUV measurements are particularly sensitive to the important stage at which the galactic fountain is being formed, since the gas is then at a higher pressure.

Although the EUV contains the majority of radiation emitted by $T \sim 10^6$ K gas, this flux is observable only for the very local region of space where $N(HI) \leq 10^{18.5}$ cm⁻². EUV background measurements (Jelinsky, Vallerga and Edelstein 1995) including preliminary results from EURD (Korpela 1998), have so far provided only upper-limits to the hot-gas emission measure that cannot sufficiently constrain models of the Galactic plasma. EUV band observations, including measurements of the Fe complex emission to be obtained by the recently announced CHIPS UNEX mission (Hurwitz, et al. 1998) can only measure the state of hot gas for the very local plasma because of IS absorption.

## The Mission

Galmatheia is the primary payload proposed for the Kitsat-4 (K4) satellite to be produced by the Korean Satellite Technology Research Center (SaTReC) for launch in 2001. The mission consists of a one-year pointed phase followed by a one-year survey of the entire sky. K-4 will dedicate its nighttime operations to Galmatheia. K-4 is a 3-axis, 1' per minute stabilized spacecraft with 1' pointing knowledge and 20' pointing accuracy.

In its first year, Galmatheia will conduct deep imaging spectroscopy of at least 300 fields, covering 150 square degrees, and will measure the primary emission lines of such important species as C III, C IV, Si IV, He II, and O IV. In a single day's observation, our mission will provide an order of magnitude improvement in both sensitivity and spectral resolution compared to the handful of existing FUV diffuse emission spectra (Martin and Bowyer 1990) upon which many current plasma cooling theories are based.

The one-year survey will chart the spatial distribution of hot Galactic plasma, producing a diffuse sky map (a hyper-spectral image) of emission from important hot plasma cooling lines such as C IV and O VI. The sky map will be carefully corrected for stellar sources and geocoronal noise. Excessive flux from FUV bright stars transiting the field of view is discarded by a fast response electronics system which preserves both exposure and mapping information accuracy while preventing an excessive burden to memory and telemetry resources.

The 5' imaging will furnish sufficient detail for productive correlation studies with other all-sky surveys that trace ISM structure, such as HI, SXR, H a, radio continuum, etc., and will allow detailed structural studies of nearby SNR shocks.

## Instrumentation

Galmatheia's bandpass, spectral resolution, sensitivity, and spatial imaging performance has been optimized for galactic ISM science. Our design solution (Edelstein, Dixon and Miller, 1998), derived from the flight-proven EURD spectrographs (Bowyer, Edelstein and Lampton 1998), provides twice the usable grasp and imaged field of conventional spectrographs. An off-axis cylindrical mirror feeds two identically figured elliptical gratings. Both spectral channels use the same optical mounting and share a slit and a planar photon-counting detector.

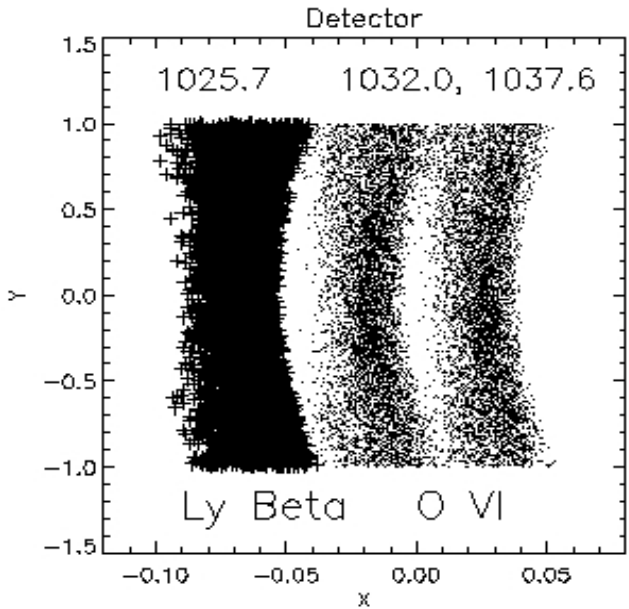

*Fig. 2 Numerical ray-trace, including the effect of both optical scattering and aberration within the spectral profile, is shown at the detector plane (scale in cm) of the Short band spectrograph. Good separation of the bright Ly b 1027 Å airglow line (crosses) from the O VI 1032/1038 Å doublet (dots) is achieved for the diffusely illuminated full field of view (8° x 5').*

We have taken great care to reduce noise from airglow lines, stars, scattered starlight continuum, and detector and ion backgrounds. The primary source of instrumental noise is instrumentally scattered geocoronal Ly a. A crystal filter excludes Ly a from the Long bandpass channel. In the Short bandpass channel, a statistically unbiased method (Bowyer, Edelstein and Lampton 1998) is used to isolate the instrumental and scattered-airglow ($l >= 1150$ Å) components of the background noise by continuously interleaving observations taken in three filtered states: un-filtered, a $MgF_2$ filter and an opaque filter. The time-varying, spatially distributed noise induced by solar and geo-magnetic interactions is thus determined by a simple arithmetic combination of the filtered spectra during ground analyses and does not require complex deconvolution or in-flight data analysis.

**Instrument Performance**

The instrument was designed using numerical ray-tracing and efficiency analyses including the effects of both optical scattering and aberration within the spectral profile (See Fig. 2). We used conservative estimates for noise sources including: geocoronal Ly a of 5 kRy to simulate non-optimum observations over the entire orbital eclipse; detector background of 2 cts/sec/$cm^2$, which is five times the observed orbital values; and a stellar continuum of 500 photon/$cm^2$/s/sr/Å (hereafter CU), which is a factor of several above reported IS values. A simulated spectrum including the modeled noise sources is shown in Fig. 3. The spectral resolution for a diffusely illuminated full field of view (8° x 5') is ~1.5 Å and ~2.3 Å HEW over almost the entire Short and Long bandpass, respectively. With a resolution of $l/dl \sim 650$, important astrophysical line profiles are distinct and the bright Ly b 1027 Å and Ly g 972 Å airglow lines are well separated from the important O VI and C III lines. The full slit HEW imaging width as determined by raytrace is 4' - 6'.

The single-emission-line, 5 s sensitivity for a one-day, 8° x 5' field of view pointing for C IV 1550 Å is 300 photon/$cm^2$/s/sr and for O VI 1032 Å is 500 photon/$cm^2$/s/sr (hereafter LU = ~ 1.5 x $10^{-11}$ ergs/sec/$cm^2$/sr = ~ 1.2 x $10^{-5}$ Rayleighs). This sensitivity yields 50 s detection of both the theoretically-predicted radiation from hot Galactic gas and previously observed levels of diffuse FUV emission. The sensitivity for different pointed integra-

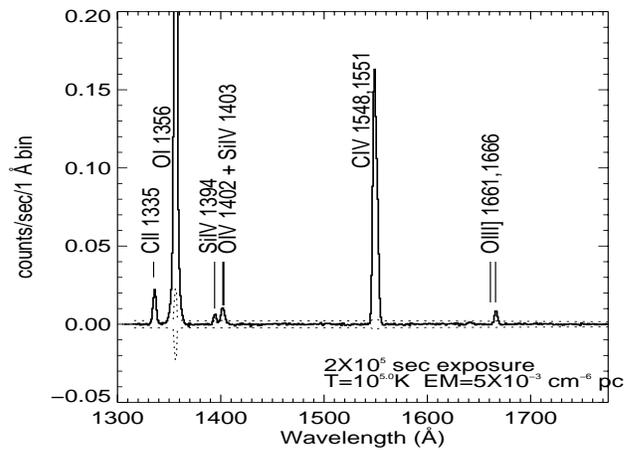

*Fig. 3 Modeled spectrum (solid line) from a one-week observation of hot ($10^5$ K) plasma with an emission measure of 0.005 $cm^{-6}$ pc. Dashed lines (barely visible) are 5 s error limits. The model includes line-profile widening from grating scatter, Poisson noise, and airglow features (e.g., O I 1356Å). Background subtraction used the multiple-filter technique described in the text. Interstellar absorption was neglected.*

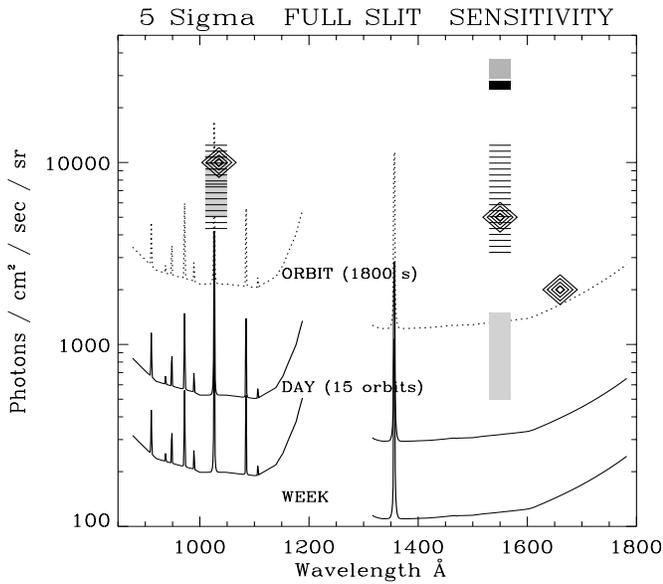

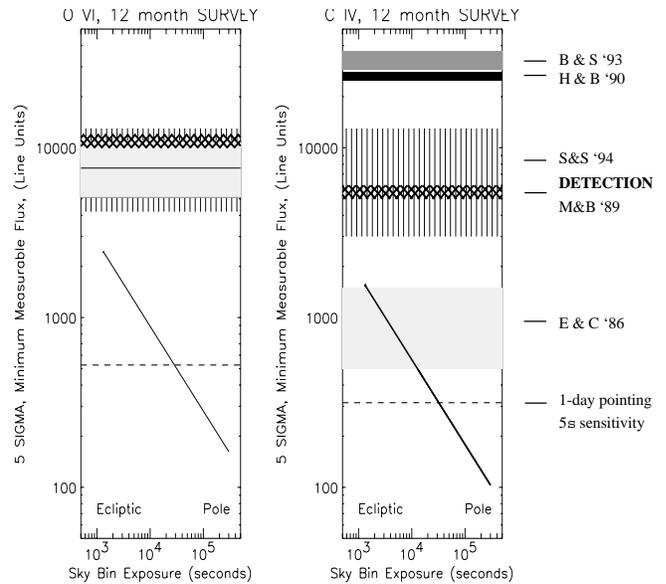

*Fig.4* Pointed mission single-emission-line, 5 s sensitivity is shown for single-orbit (dashed line), -day (central solid line), and -week (lower solid line) integrations of the 8° x 5' field of view. Peaks in the curves correspond to airglow lines. Previous detections of O VI 1035Å, C IV 1550Å, and OIII 1663Å are plotted as diamonds. Model predictions are represented by vertical bars using the same symbols as Fig. 5.

*Fig 5* Sky-map single-emission-line 5 s sensitivity (thick dashed line) for OVI 1032 Å (left) and CIV 1550 Å (right) as a function of the integrated exposure time in a 3° x 3° sky bin after a one-year survey. Previous detections and model predictions are overplotted (shaded areas). For comparison, the pointed mission, single-day integration, 8° x 5' field sensitivity is show

tion times is shown in Fig.4.

The long-wavelength-band continuum sensitivity for a one-day pointing is 30 CU at 5 s significance over the 8° x 5' field of view, three to five times fainter than the best existing few-s significance continuum limits. This sensitivity provides 15-25 s detection of the flux predicted to arise from unresolved extra-galactic sources (Bowyer 1989).

The sky-survey map median sensitivity for a 3° x 3° bin of the sky survey is 1000 LU for C IV 1550 Å and 1500 LU for O VI 1032 Å. The survey sensitivity will allow 10-15 s detection of both the theoretically-predicted flux from hot Galactic gas and previously-observed levels of FUV emission. Models of the Galactic corona will be readily distinguished. The map can be binned into larger sky pixels for improved sensitivity. The sensitivity may also be increased by reducing the fraction of sky coverage elected during operations. The survey sensitivity is shown in Fig. 5.

## Conclusion

Galmatheia will measure the energetics of hot IS plasma that shapes the Galaxy and drives its evolution on local to coronal scales. These measurements will enhance our understanding of the structure, mixing, and cooling of hot IS plasmas by testing multi-phase models of the ISM, examining the evolution of hot gas in superbubbles, and observing shock interactions between SNe and the ISM. Our data will complement the thermal diagnostics from current and future missions (e.g., STIS, FUSE, CHIPS, ROSAT) and ground-based facilities and, combined with absorption-line data from specific lines of sight, will allow the unique determination of important physical parameters of the hot ISM.

**Acknowledgements**

The authors acknowledge the important contributions of T.N. Miller, J. and P. McCauley, and E. Sokolov. This work was supported by the Space Sciences Laboratory .